\documentclass[twocolumn,showpacs,showkeys,preprintnumbers,amsmath,amssymb]{revtex4}

\usepackage{graphicx}% Include figure files
\usepackage{dcolumn}% Align table columns on decimal point
\usepackage{bm}% bold math

\begin{document}
\title{Stellar ($n, \gamma$) cross sections of $^{23}$Na}
\author{E. Uberseder}
\affiliation{Karlsruhe Institute of Technology (KIT), Campus North, Institute of Nuclear Physics, PO Box 3640, Karlsruhe, Germany}
\affiliation{University of Notre Dame, Department of Physics, Notre Dame, IN 46556, USA}
\altaffiliation[present address:]{Texas A\&M University, TX 77843, USA} 
\author{M. Heil}
\email[corresponding author: ]{m.heil@gsi.de}\affiliation{ GSI Darmstadt, Planckstr. 1, 64291 Darmstadt, Germany}
\author{F. K\"appeler}
\affiliation{Karlsruhe Institute of Technology (KIT), Campus North, Institute of Nuclear Physics, PO Box 3640, Karlsruhe, Germany}
\author{C. Lederer}  
\affiliation{School of Physics and Astronomy, University of Edinburgh, UK}
\author{A. Mengoni}
\affiliation{ENEA, 40100 Bologna, Italy}
\author{S. Bisterzo} \affiliation{ INAF, Astrophysical Observatory of Turin, 10025 Pino Torinese, Italy}
\affiliation{INFN, Sezione di Torino, Via P. Giuria 1, 10125 Torino, Italy}
\author{M. Pignatari} \affiliation{E.A. Milne Centre for Astrophysics, Dept of Physics \& Mathematics, University of Hull, HU6 7RX, United Kingdom}
\affiliation{NuGrid collaboration, \url{http://www.nugridstars.org}}
\author{M. Wiescher}
\affiliation{University of Notre Dame, Department of Physics, Notre Dame, IN 46556, USA}

\date{\today}

\begin{abstract}
The cross section of the $^{23}$Na($n, \gamma$)$^{24}$Na reaction has been
measured via the activation method at the Karlsruhe 3.7 MV Van de Graaff
accelerator.  NaCl samples were exposed to quasistellar neutron spectra at
$kT=5.1$ and 25 keV produced via the $^{18}$O($p, n$)$^{18}$F 
and $^{7}$Li($p, n$)$^{7}$Be reactions, respectively.  The derived
capture cross sections 
$\langle\sigma\rangle_{\rm kT=5 keV}=9.1\pm0.3$ mb and
$\langle\sigma\rangle_{\rm kT=25 keV}=2.03 \pm 0.05$ mb are
significantly lower than reported in literature. These results were used to 
substantially revise the radiative width of the first $^{23}$Na resonance and to 
establish an improved set of Maxwellian average cross sections. The
implications of the lower capture cross section for current models of $s$-process
nucleosynthesis are discussed.
\end{abstract}



\maketitle

\section{Introduction}

Sodium represents a neutron poison for the slow neutron capture ($s$) process,
particularly in massive stars with more than about eight solar masses ($>$8 M$_\odot$) \cite{RGB93}. The $s$ process in massive stars is particularly efficient in producing species in the mass range 60$<$A$<$90, forming the weak $s$-process component in the inventory of the solar abundances. In addition to its importance for the neutron balance in massive stars, the
($n, \gamma$) cross section of $^{23}$Na is also needed to follow the production of 
sodium in low and intermediate mass asymptotic giant branch (AGB) stars. In these stars, the 
$main$ $s$ process contributes most of the $s$ abundances in the solar system from Zr to 
Pb, and the $strong$ $s$-process adds to the Pb/Bi abundances at the termination point 
of the $s$-process reaction path \cite{GAB98,KGB11}.

The weak $s$ process in massive stars begins at the end of convective core He-burning 
($T_8>2.5$), where $^{22}$Ne($\alpha, n$)$^{25}$Mg operates as the principal neutron 
source. During that period, sodium is only marginally produced by neutron captures on 
$^{22}$Ne. During the subsequent convective C-shell burning, sodium is efficiently made
via the $^{12}$C($^{12}{\rm C}, p$)$^{23}$Na channel although most of the protons 
(and sodium) are consumed by $^{23}$Na($p, \alpha$)$^{20}$Ne reactions \cite{ThA85}.
Nevertheless, the C-burning layers of massive stars ejected in the subsequent supernova (SN)
are one of the major sources of sodium in the Galaxy \cite{WHW02,KUN06}, together with stellar 
winds from AGB stars (e.g. \cite{LMJ11}). In the convective C-burning shell, neutrons are
mainly released via $^{22}$Ne($\alpha, n$)$^{25}$Mg reactions as $^{22}$Ne is present in
the ashes of the convective He-burning core and $\alpha$ particles are liberated in
$^{12}$C($^{12}{\rm C}, \alpha$)$^{20}$Ne reactions (e.g. \cite{RBG91a}). 

In the weak $s$ process most of the neutrons are captured by abundant light isotopes,
which act as neutron poisons, and only a small fraction is available for captures on iron 
seed nuclei to feed heavy isotope nucleosynthesis.  At solar metallicity, more 
than 70\% of the available neutrons are captured by neutron poisons in the He-burning 
core, and more than 90\% in the C-burning shell.  For this reason, it is extremely important 
to quantify the neutron capture rates of light isotopes such as sodium to evaluate 
the impact of neutron poisons in the weak $s$ process.  

Another relevant source for the production of sodium are thermally pulsing low-mass 
(e.g., \cite{Mow99, GoM00, CPS11}) and massive AGB stars \cite{Kar10}, where the 
$s$ process is related to the He shell burning stage of evolution. 
In a first episode, neutrons are produced by $^{13}$C($\alpha, n$)$^{16}$O reactions 
during the interpulse phase between He shell flashes at temperatures of $T_8$=0.9 
($kT$=8 keV) \cite{BGW99}. The mixing of protons with the top layer of the He shell, 
required to provide the necessary $^{13}$C for neutron production, has the additional 
effect of activating the NeNa cycle in the partial mixing zone \cite{GoM00}, which then 
continues in the H-burning shell throughout the interpulse phase \cite{Mow99}. A second, 
weaker neutron exposure takes place during the He shell flash at higher temperatures of 
$T_{8}$=2.6 ($kT$=23 keV) when the $^{22}$Ne($\alpha, n$)$^{25}$Mg source is 
marginally activated. As the He flash engulfs the ashes of the H burning shell, further 
$^{23}$Na might be produced by neutron captures on the abundant $^{22}$Ne during 
this second phase of $s$-processing in thermally pulsing AGB stars. Recent studies by 
Cristallo {\it et al.} \cite{CGS06} and Bisterzo {\it et al.} \cite{BGS10} confirm that 
neutron capture production of primary sodium is particularly efficient in low-mass AGB 
stars of low metallicity.   

Intermediate-mass AGB models experience hot hydrogen burning (HBB), which
modifies the Na abundance on the stellar surface depending on the attained
temperature and on the interplay with the efficiency for third dredge up 
\cite{VeD11,SCP14}.

Despite of its relevance for nuclear astrophysics, the Maxwellian averaged neutron 
capture cross section (MACS) of $^{23}$Na is rather uncertain \cite{MAM78b}. In 
this work, we present new experimental data for $^{23}$Na measured in quasi-stellar 
neutron spectra at thermal energies of $kT=5.1$ and 25 keV. Appropriate spectra have 
been obtained via the $^{18}$O($p, n$)$^{18}$F and $^{7}$Li($p, n$)$^{7}$Be 
reactions to simulate stellar temperature conditions relevant to $s$-process 
nucleosynthesis. The experimental details and results of the activation measurements 
are given in Secs. \ref{expsection} and \ref{anasection}. In Sec. \ref{implications}, 
MACS values are derived for the full range of $s$-process temperatures on 
the basis of the present results. The implications of these data for the $s$-process 
abundances are discussed for massive stars as well as for AGB stars.

\section{Experiment \label{expsection}}

Similar to many light nuclei, the $^{23}$Na($n, \gamma$)$^{24}$Na cross section 
is difficult to study experimentally given the high ratio of scattering to capture cross 
sections. In such cases, neutrons scattered on the sample and subsequently captured 
in or near the detector can induce a large background when measuring prompt capture 
gammas with the time of flight (TOF) technique \cite{KWG00}. These difficulties can 
be avoided with the activation method, because the induced activity of the product 
nucleus is counted only after the irradiation in a low background environment. Therefore, 
the activation technique is well suited to measure ($n,\gamma$) cross sections of light
nuclei with greater precision than reported previously from TOF measurements.

The experiment was carried out by a series of repeated irradiations with a set of
different samples and by variation of the relevant activation parameters. In this 
way, corrections concerning the dimensions of the samples, self absorption, 
and the decay during irradiations could be constrained and the determination 
of systematic uncertainties improved. 

\subsection{Samples}

The samples for the individual runs were  prepared from NaCl (99.99\%
pure) pressed into cylindrical pellets 6, 8, 10, and 15 mm in diameter with
varying thicknesses. As NaCl is hygroscopic, care was taken to be sure that
the water content of the material gave a negligible contribution to the
mass. This was verified by heating a quantity of the NaCl at 250 $^\circ$C 
for two hours, showing  that the sample mass before and after heating differed 
by less than 0.01\%. 

The sodium cross section was measured relative to that of gold, which is
commonly used as a reference in activation measurements. Gold foils 0.03 mm
in thickness were cut to the proper diameters and fixed to the front and back of
the samples during irradiation. The sample masses, as well as those of the 
respective gold foils, are given in Table \ref{tab1}.

\begin{table*}
\caption{Sample characteristics\label{tab1}}
\begin{ruledtabular}
\begin{tabular}{lccccc}
Sample 	& Diameter (mm)	& \multicolumn{2}{c}{Sample mass$^a$ (mg)} 	& \multicolumn{2}{c}{Mass of Au foils (mg)}\\ 
\cline{3-4}\cline{5-6}
          	&	 		& NaCl 				& Na  		& upstream 		&  downstream     	\\
\hline
Na-1	&  6 		& 54.06 				& 21.27 		& 16.49 			& 16.40				\\
Na-2	&  8 		& 82.67 				& 32.52 		& 27.92 			& 28.47				\\
Na-3	&  10 		& 88.28 				& 34.73 		& 45.36 			& 43.54				\\
     		&			&         				&           	&          			&          				\\ 
Na-4	&  10		& 232.4 				& 91.42 		& 43.92 			& 44.07				\\
Na-5	&  10 		& 307.6 				& 121.0 		& 43.80 			& 42.98				\\
Na-6	&  15 		& 500.1 				& 196.7 		& 97.76 			& 98.47				\\
\end{tabular}
\end{ruledtabular}
\footnotetext[1]{Uncertainties always less than 0.05 mg.}\\
\end{table*}

\subsection{Neutron irradiations}

All neutron irradiations were performed at the Karlsruhe 3.7 MV Van de Graaff
accelerator. The $^7$Li($p, n$)$^7$Be reaction was utilized to obtain a 
thermal neutron spectrum at $kT=25$ keV \cite{BeK80,RaK88,LKM12,FFK12}. 
The target was produced by evaporation of 30 $\mu$m of lithium onto a 
water-cooled copper backing and was transported to the beamline in an argon 
atmosphere to prevent oxidation. The required proton beam energy of 
$E_p=1912$ keV was adjusted 31 keV above the reaction threshold at 1881 keV. 
Due to the kinematics of the $^7$Li($p,n$)$^7$Be reaction, neutrons are 
emitted in a forward cone with an opening angle of 120$^\circ$. As the 
Maxwellian neutron distribution is obtained by angular integration over the 
entire forward cone, care was taken to be sure that the samples of varying 
diameters always extended through the full solid angle. 

\begin{figure*}
\includegraphics[width=2.5in]{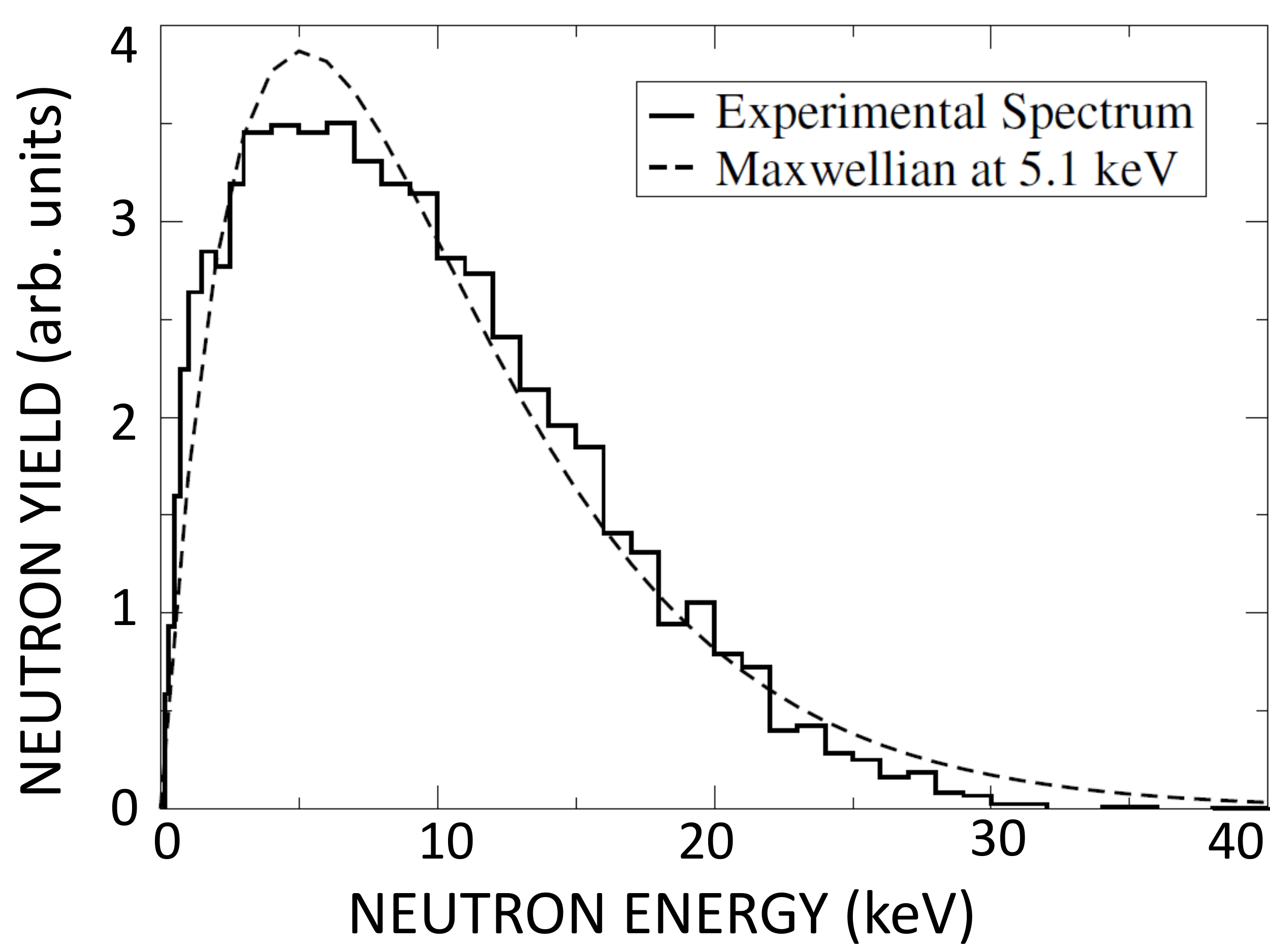}
\hspace*{15mm}
\includegraphics[width=2.5in]{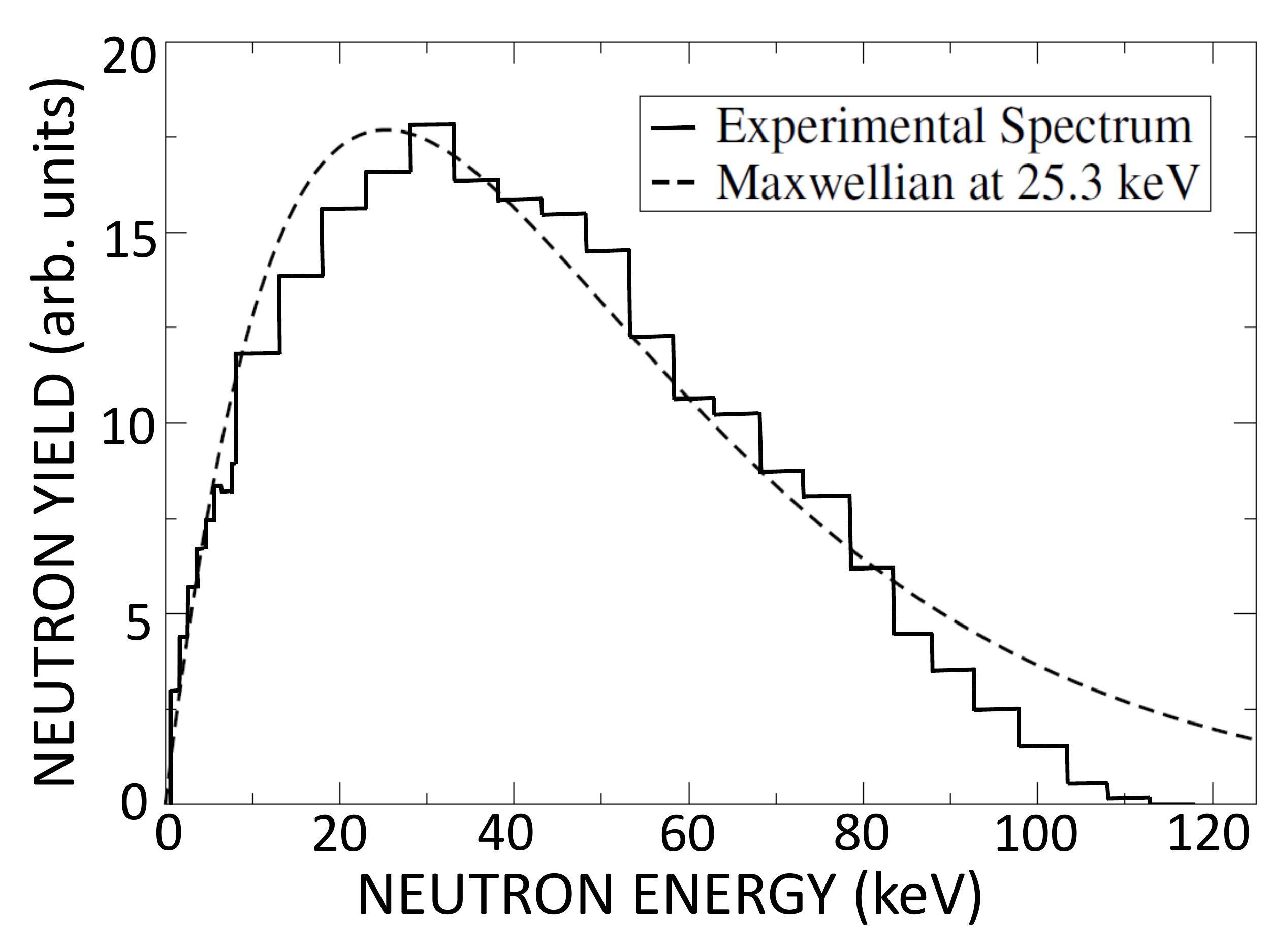}
\caption{Experimental neutron distributions at $kT=5.1$ keV (left) and 25 keV (right)
compared to the respective Maxwell-Botzmann spectra. \label{fig1}}
\end{figure*}

Correspondingly, the $^{18}$O($p, n$)$^{18}$F reaction has been used to 
reproduce a thermal neutron distribution at $kT =5.1$ keV \cite{HDJ05} as 
shown in Fig. \ref{fig1}. The target consisted of a 0.2-mm-thick tantalum 
disk with an oxide layer about 2 $\mu$m in thickness, which was produced
by electrolysis of water with 95\% enrichment in $^{18}$O. The
tantalum disk was glued onto a 1-mm-thick water-cooled copper backing. 
The glue was suited for vacuum applications and was particularly selected
to obtain good heat contact with the copper backing. Also in this case, the 
proper proton energy of $E_p=2582$ keV was calibrated against the nearby 
reaction threshold at $E_p = 2574$ keV, and care was taken that the sample 
extended through the full forward cone of the neutron beam.

The neutron flux during the runs at $kT=25$ keV 
was recorded by a lithium glass monitor, mounted on the beam axis at a 
distance of 831 mm from the target. Due to the much lower neutron yield for 
the irradiations at $kT=5.1$ keV, the lithium glass monitor was replaced by a 
more sensitive ZnS detector. The irradiation times, $t_b$, for each run are listed 
in Table \ref{tab2}. The proton beam current was typically 100 $\mu$A
in all irradiations.

\begin{table}
\caption{Irradiation, transfer, and measuring times \label{tab2}}
\begin{ruledtabular}
\begin{tabular}{cccc}
Sample 	& $t_b$ 		& $t_w$ 	& $t_m$ \\
 		& (h) 		& (min) 		& (h) 	\\
\hline
Na-1 	& 16.3 		& 34.7 		& 8.56	\\
Na-2	& 21.0 		& 26.8 		& 10.0	\\
Na-3 	& 24.7 		& 33.0 		& 5.00	\\
     		&        		&        		&        	\\
Na-4 	& 13.8 		& 7.20 		& 72.7	\\
Na-5 	& 17.6 		& 3.00 		& 21.6	\\
Na-6 	& 19.6 		& 4.00 		& 22.2	\\
\end{tabular}
\end{ruledtabular}
\end{table}

\subsection{Measurement of induced activity}

A high purity (HP) Ge detector of 39 cm$^3$ was utilized to measure the 
induced activities in the sodium samples and gold foils following the
$kT = 25$ keV runs. The efficiency of the detector as a function of energy was
determined with several calibrated sources to an accuracy of 1.5\%. A relatively
large source-detector distance of 76 mm reduced the effect of coincidence
summing. To reduce natural backgrounds, the detector was shielded with 50
mm lead and 5 mm copper. The measuring times, $t_m$, as well as the 
waiting times between irradiation and the $\gamma$ counting, $t_w$, are 
listed in Table \ref{tab2}. 

\begin{figure}
\includegraphics[width=2.5in]{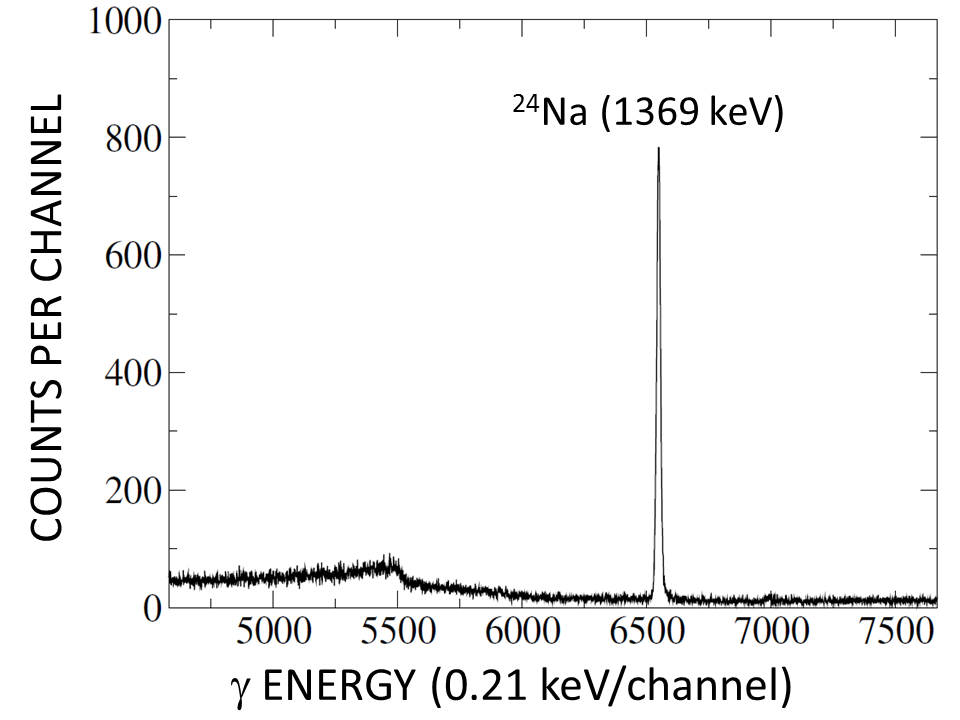}
\caption{$\gamma$ Spectrum for sample Na-1 after irradiation at $kT=25$ keV. \label{fig2}}
\end{figure}

Due to the much lower neutron flux reached during the irradiations with the 
$^{18}$O target, the weak induced activities had to be determined with a 
more sensitive setup consisting of two HPGe clover detectors with the sample 
centered between the detectors in very close geometry as described in detail 
by Dababneh {\it et al.} \cite{DPA04}.  Again, the photo-peak efficiency was 
determined using several calibrated sources. In this case, the small source-detector 
distance required the application of corrections for summing effects and for the
sample size. These, along with self absorption corrections, were determined 
by Monte Carlo simulations with the GEANT4 toolkit \cite{Gea03} using a 
precise modeling of the apparatus. The 1369 keV $\gamma$-ray from 
the $^{24}$Na decay was chosen for the analysis and appeared prominently 
above background. Figure \ref{fig2} shows the 1369 keV line from 
the 6 mm NaCl sample after activation in the $kT=25$ keV spectrum. It should 
be noted that a short-lived (20 ms) isomeric state exists in $^{24}$Na, which 
has a very small probability to decay directly to $^{24}$Mg without producing 
a 1369 keV $\gamma$-ray.  Endt {\it et al.} \cite{EBF98} quote this probability 
to be 0.05\%, thus it was considered to have a negligible effect on the present 
measurement.  

\section{Analysis and Results\label{anasection}}

The process to extract a cross section from the activation data follows 
the approach of Ref. \cite{BeK80}.  The number of activated nuclei is defined as 
\begin{equation} 
\label{activity} 
A = \Phi_{tot} N \sigma f_b,
\end{equation} 
\noindent
where $\Phi_{tot} = \int \Phi(t) dt$ represents the time integrated neutron
flux, $N$ is the number of sample atoms, and $\sigma$ the cross section in 
cm$^2$. The correction 
\begin{displaymath} 
f_b=\frac{\int^{t_b}_0 \Phi(t) e^{-\lambda(t_b-t)}dt}{\int^{t_b}_0 \Phi(t)}dt.
\end{displaymath} 
for decays of the produced nuclei during the irradiation phase was determined 
from the time-dependent yield recorded by the before mentioned lithium glass or zinc 
sulfide monitors, which was recorded as counts per 60 s intervals.  The time 
integrated neutron flux was determined from the activity of the gold foils. 
The counts registered in the HPGe detectors are expressed as a function of the 
induced activity,
\begin{equation} 
\label{counts} 
C = A  K_\gamma \epsilon_\gamma I_\gamma e^{-\lambda t_w} (1-e^{-\lambda t_m}). 
\end{equation}  

\noindent
where $K_\gamma$ is the total correction factor for self absorption, summing, 
and sample size effects, $\epsilon_\gamma$ the peak efficiency of the HPGe 
detector, and $I_\gamma$ the absolute gamma intensity per decay of the 
activated nuclei. These quantities are given in Table \ref{tab3}. 

\begingroup
\begin{table*}
\caption{Decay characteristics and absolute $\gamma$ efficiencies \label{tab3}}
\begin{ruledtabular}
\begin{tabular}{cccccc}
Nucleus & t$_{1/2}$ & $E_\gamma$ (keV) & $I_\gamma$ (\%) & 
  \multicolumn{2}{c}{$\epsilon_\gamma$ (\%)} \\
\cline{5-6} 
& & & & $kT = 5.1$ keV & 25 keV \\
\hline
$^{24}$Na &  14.9590$\pm$0.0012 h& 1368.63 & 100.00$\pm$0.0\footnotemark[1] & 6.94$\pm$0.10 & 0.048$\pm$0.001 \\
$^{198}$Au &2.69517$\pm$0.00021 d& 411.80 & 95.58$\pm$0.12\footnotemark[2] & 20.6$\pm$0.3 & 0.224$\pm$0.01 \\
\end{tabular}
\end{ruledtabular}
\footnotetext[1]{Ref. \cite{EBF98}}
\footnotetext[2]{Ref. \cite{Chu02}}
\end{table*}
\endgroup

The adopted half-lives of $^{24}$Na and $^{198}$Au were 14.9590$\pm$0.0012 h 
\cite{EBF98} and 2.69517$\pm$0.00021 d \cite{Chu02}, respectively.  

The $^{197}$Au cross sections  that are needed for 
the determination of $\Phi_{tot}$ were determined in the following way: The cross sections 
averaged over the experimental spectra as well as the respective Maxwellian average spectra 
at $kT=5$ and 25 keV were calculated using the evaluated gold cross section from the 
JEFF-3.2 data library \cite{JEF14}. These results were then corrected for the 1.1 and 1.2\% differences 
found with respect to the recommended MACS values in the KADoNiS-1.0 compilation
\cite{DPK14}. The adopted spectrum-averaged gold cross sections are $\sigma_{exp} = 
637\pm13$ mb for $kT=25$ keV and $\sigma_{exp} = 1922\pm70$ mb for $kT=5.1$ keV.   

\begingroup
\begin{table}
\caption{Measured ($n, \gamma$) cross sections of $^{23}$Na (in mb)\label{tab4}}
\begin{ruledtabular}
\begin{tabular}{lclc}
\multicolumn{2}{c}{$kT=5.1$ keV} & \multicolumn{2}{c}{$kT=25$ keV} \\ 
\cline{1-2} \cline{3-4}
Sample	& $\sigma_{exp}$$\pm$syst$\pm$stat	& Sample & $\sigma_{exp}$$\pm$syst$\pm$stat\\ 
\hline
Na-4 	& 9.03$\pm$0.28$\pm$0.26 			& Na-1 & 2.05$\pm$0.06$\pm$0.02	\\
Na-5 	& 8.94$\pm$0.27$\pm$0.14 			& Na-2 & 2.02$\pm$0.05$\pm$0.01 	\\
Na-6 	& 9.33$\pm$0.29$\pm$0.14 			& Na-3 & 2.03$\pm$0.05$\pm$0.02 	\\
\\ 
Average & 9.10$\pm$0.28$\pm$0.11 			& & 2.03$\pm$0.05$\pm$0.01 		\\
\end{tabular}
\end{ruledtabular}
\end{table}
\endgroup

The experimental results of the activations are summarized in Table
\ref{tab4}. Statistical and systematic errors are quoted separately. The
statistical errors arise from the counting of the induced activities and from
the corrections applied as a result of the Monte Carlo simulations. Sources of
systematic error are listed in Table \ref{tab5}.  

\begingroup
%\squeezetable
\begin{table}
\caption{Systematic uncertainties (in \%) of the measured ($n, \gamma$) cross 
           section of $^{23}$Na\label{tab5}}
\begin{ruledtabular}
\begin{tabular}{lcccc}
Source & \multicolumn{2}{c}{$kT=5.1$ keV} & \multicolumn{2}{c}{$kT=25$ keV}\\
\cline{2-3} \cline{4-5}
 									& Au 	& Na 	& Au 	& Na	\\
\hline
Time and decay factors 				& 0.5	& 0.5	& 0.6 	& 0.5	\\
Au cross section 						& 2.0 	& N/A 	& 1.4 	& N/A 	\\
Sample atoms 						& 0.1 	&$<$0.1& 0.2 	& 0.1	\\
Self absorption 						& N/A 	& N/A 	&$<$0.1&$<$0.1\\
Detector peak efficiency				& 1.5 	& 1.5 	& 1.5 	& 1.5	\\
Intensity of $\gamma$-decay branch	& 0.1 	&$<$0.1& 0.1 	&$<$0.1\\
Neutron flux $\Phi_T$ 				& N/A 	& 3.8  	& N/A 	& 2.1	\\
									&		&		&		&		\\
Total  								& 		& 3.1 	& 		& 2.7	\\
\end{tabular}
\end{ruledtabular}
\end{table}
\endgroup

\section{Astrophysical Implications\label{implications}}

\subsection{Maxwellian Averaged Cross Sections\label{macs}}

As neutrons are quickly thermalized in a stellar environment, the stellar 
spectra can be described by a Maxwell-Boltzmann distribution. The MACS 
of an isotope exposed to such a neutron distribution is   
\begin{equation} 
\label{eq:macs} 
\langle\sigma\rangle_{kT} = \frac{\langle\sigma v\rangle}{v_T} = \frac{2} {\sqrt{\pi}} \frac{\int^\infty_0 \sigma(E_n) E_n e^{-E_n/kT} dE_n}{\int^\infty_0 E_n e^{-E_n/kT}dE_n}  
\end{equation} 
\cite{BBK00}. 
While the experimental spectra closely reproduce Maxwellian distributions, the
small differences from a true MACS must still  be accounted for. This is 
done by folding the experimental neutron distribution with the capture cross
section provided by a database to yield a normalization, which is then applied
to produce a true MACS calculated from the evaluated energy-dependent cross section.
This method supplies a sufficiently accurate approximation for the MACS in the
local energy region of the measurement, but is prone to increasing uncertainty 
in extrapolation to cover the full thermal energy region of stellar $s$-process 
scenarios from $kT\approx$8-90 keV. 

The MACS values for $^{23}$Na at $kT= 5$ and 25 keV are influenced by a
broad s-wave resonance at 2.8 keV, a smaller p-wave resonance at 35 keV, and a
broad p-wave resonance at 53 keV.  For the present analysis, the ($n, \gamma$) 
cross section was modeled by means of the R-matrix code SAMMY \cite{Lar06}, 
starting with resonance energies and widths of all resonances up to 500 keV listed 
in the ENDF/B-VII.1 data library \cite{CHO11}. 

In addition to the resonance terms, an s-wave direct radiative capture (DRC) 
component in the $^{23}$Na cross section was included by a set of optical 
model DRC calculations \cite{MOI95}, using the spectroscopic factors of Ref. 
\cite{TNB04}. The contribution of the DRC component to the measured 
spectrum-averaged cross section is rather small but it is important for obtaining 
a consistent reproduction of the thermal cross section data. In fact, a DRC 
component of 170 mb at thermal energy with a 1/v energy dependence is 
required to reproduce the thermal capture cross section as well as to match 
the resonance integral. 

The resulting energy-dependent cross section was then folded with the experimental 
neutron spectra to produce cross sections that could be directly compared with the 
activation results. It turned out that it was sufficient to reduce only the
radiation width of the resonance at 2.8 keV by 35\% in order to reproduce
the measured cross sections, while leaving the widths of the other resonances 
fixed. As shown in Table \ref{tab:sam}, the measured cross sections and the 
corresponding values obtained with this best-fit solution agreed to better than 
2\%. Without any further changes, the thermal cross section \cite{FRB14} and 
also the resonance integral \cite{Mug06} could be reproduced by this best fit 
as well. 

The uncertainty of the radiative width for the 2.8 keV resonance was determined
by variation of the SAMMY fit to match the uncertainty of the measured cross 
section values, yielding $\Gamma_\gamma = 245\pm8$ meV. This corresponds 
to an improvement by a factor of three compared to the (optimistic) 10\% 
uncertainty quoted in the early work by Friesenhahn {\it et al.} \cite{FLF68}. 

\begin{table*} [tb]
\caption{Measured cross sections and calculated spectrum-averaged values  
\label{tab:sam}}
\begin{ruledtabular}
\begin{tabular}{lccc}
Neutron spectrum  	& \multicolumn{3}{c}{Spectrum-averaged cross sections (mb)}	\\
\cline{2-4}
(keV)					& ENDF/B-VII.1	& Experimental value	&  SAMMY best-fit\\
\hline
25.3 10$^{-6}$ (thermal)	& 528			& 541$\pm$3$^a$	& 542			\\
Resonance integral		& 312.3			& 311$\pm$10$^b$	& 300			\\
5.1 (q-MB)				& 12.5			& 9.10$\pm$0.30		& 9.07			\\		
25 (q-MB)				& 2.16			& 2.03$\pm$0.05		& 2.00			\\
\end{tabular}
\end{ruledtabular}
\footnotetext[1]{Ref. \cite{FRB14}}\\
\footnotetext[2]{Ref. \cite{Mug06}}
\end{table*}

With the corrected strength of the 2.8-keV resonance, the best-fit solution for
the capture cross section has been used to establish the updated set of MACS
values for $^{23}$Na. The effect of the reduced and more accurate 
$\Gamma_\gamma$ value of the 2.8 keV resonance is illustrated in Table 
\ref{tab6} by the comparison of these results with the data set in the KADoNiS 
v0.3 compilation \cite{DPK09} and with the evaluated cross section from the 
ENDF/B-VII.1 data library \cite{CHO11}, which shows a significant reduction 
in the MACS values especially at thermal energies below 30 keV. The effect of 
the revised radiative width of the 2.8 keV resonance is most pronounced for 
the MACS values at $kT=5$ and 10 keV, which are lower by $\sim30$ and 20\%, 
respectively, whereas it reduces the 25-keV MACS only by $\sim4$\%.

The uncertainties quoted for thermal 
energies of $kT=5$ and 25 keV could directly be taken from the experimental 
values, but slightly larger uncertainties have to be considered otherwise to 
account for the different weights of the involved resonances, which are changing 
with thermal energy. Below 30 keV this additional uncertainty is less than 
3\%, because only the first resonance had to be corrected to reproduce the 
25-keV cross section. Towards higher values of $kT$, where an increasing 
number of resonances is contributing, this uncertainty is expected to grow 
and may reach 10\% at $kT=100$ keV.

\begin{figure}
\includegraphics[width=3in]{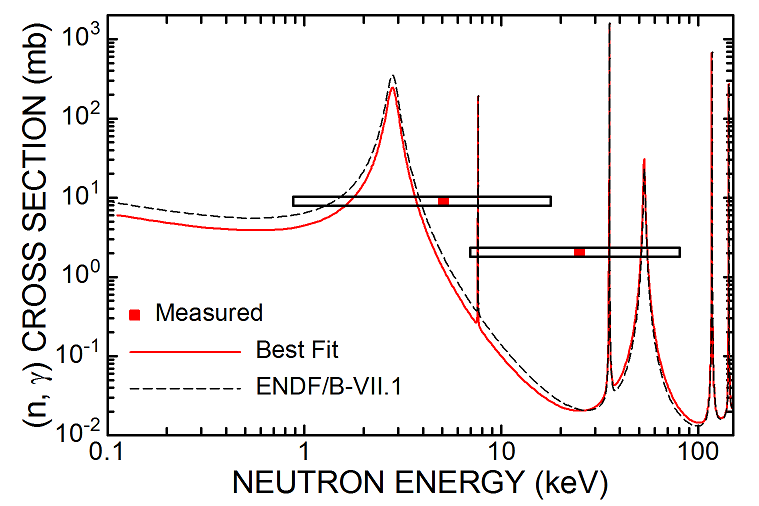}
\caption{The measured spectrum averaged cross sections (full squares) 
and the best-fit solution obtained with the R-matrix code SAMMY 
\cite{Lar06}, which is fully consistent with the data points (solid line). 
The FWHM of the neutron spectra are indicated by open boxes. 
The comparison with the ENDF/B-VII.1 evaluation \cite{CHO11} (dashed 
line) shows that it was sufficient to correct only the first resonance at 
2.8 keV. \label{fig:comp}}
\end{figure}
 
\begingroup
\begin{table*} [tb]
\caption{Maxwellian averaged cross sections (mb) of the  
            $^{23}$Na($n, \gamma$)$^{24}$Na reaction \label{tab6}}
\begin{ruledtabular}
\begin{tabular}{lccccccccccc}
$kT$ (keV) & 5 & 10 & 15 & 20 & 25 & 30 & 40 & 50 & 60 & 80 & 100\\
\hline
This work 			& {\bf 10.1 $\pm$ 0.4} 	& 4.12 	& 2.83 	& 2.30 	& {\bf 2.00 $\pm$ 0.05} 	& 1.80 			& 1.55 	
					& 1.40 					& 1.30 	& 1.16 	& 1.05\\
KADoNiS-0.3$^a$ 	& 14  					&  5.2	& 3.4 	&  2.7	&          2.2        			& 2.1$\pm$ 0.2	& 1.7 								&  1.5					& 1.4   	&  1.3  	& 1.2 \\
ENDF/B-VII.1 \cite{DPK14} & 13.9 				& 5.24 	& 3.29 	& 2.51	& 2.09 					& 1.83$\pm$0.2      & 1.52	 						& 1.35 					& 1.24 	& 1.09 	& 0.99\\
\end{tabular}
\end{ruledtabular}
\footnotetext[1]{Ref. \cite{DPK09}}
\end{table*}
\endgroup

The correction of these MACS values for the effect of thermally excited nuclear states, 
the so-called stellar enhancement factor, is negligible over the entire range of $s$-process 
temperatures \cite{RMD11}.

\subsection{Massive stars}

The $s$ process in massive stars is known to produce most of the 
$s$ isotopes in the solar system between Fe and Sr (see \cite{KGB11} and references 
therein). In the convective He core, the neutron exposure starts to increase only in the last 
phase, close to He exhaustion, when the temperature is high enough to efficiently burn 
$^{22}$Ne via $^{22}$Ne($\alpha, n$)$^{25}$Mg. The $^{22}$Ne 
available at the end of the He core phase is given by the initial abundance of the CNO nuclei.  
As CNO elements are converted to $^{14}$N in the previous H-burning core, $^{14}$N is 
converted to $^{18}$O via the reaction channel 
$^{14}$N($\alpha, \gamma$)$^{18}$F($\beta^+$)$^{18}$O at the beginning of the 
He-burning core and then to $^{22}$Ne by $\alpha$-captures when the temperature exceeds 
$T_8$ = 2.5.  At the point of He exhaustion the most abundant isotopes are $^{16}$O, $^{12}$C, 
$^{20,22}$Ne and $^{25,26}$Mg, where the final $^{12}$C and $^{16}$O abundances 
are defined by the $^{12}$C($\alpha$,$\gamma$)$^{16}$O reaction. In He core conditions, 
$^{23}$Na is produced by the neutron capture channel 
$^{22}$Ne($n, \gamma$)$^{23}$Ne($\beta^-$)$^{23}$Na, and it is depleted via 
$^{23}$Na($n, \gamma$)$^{24}$Na.

In the convective C shell the neutron exposure starts to increase during C ignition at the 
bottom of the shell, where neutrons are mainly produced again by the $^{22}$Ne($\alpha, 
n$)$^{25}$Mg reaction. Typical temperatures at the bottom of the C shell are T$\approx$ 
1 GK, almost constant during the major part of the shell development (e.g., 
\cite{TEM07, PGH10}). In the last day(s) before the SN, temperatures 
at the base of the C shell may increase due to  thermal instabilities in the deeper 
O-burning layers, and if the C shell is still fully convective, C-shell nucleosynthesis will 
be revived \cite{PGH10}. At the end of the convective C-burning shell the most 
abundant isotopes are $^{16}$O, $^{20}$Ne, $^{23}$Na and $^{24}$Mg. Sodium 
is mainly produced via the C-burning reaction $^{12}$C($^{12}$C, $p$)$^{23}$Na 
and marginally via $^{22}$Ne($p, \gamma$)$^{23}$Na and 
$^{22}$Ne($n, \gamma$)$^{23}$Ne($\beta^-$)$^{23}$Na. The strongest sodium 
depletion reaction is $^{23}$Na($p, \alpha$)$^{20}$Ne, with smaller contributions 
from $^{23}$Na($p, \gamma$)$^{24}$Mg and $^{23}$Na($n, \gamma$)$^{24}$Na.

The impact of the new $^{23}$Na($n, \gamma$)$^{24}$Na cross section on the weak 
$s$-process distribution was studied with the NuGrid post-processing code MPPNP 
\cite{HDF08} for a full 25 M$_{\odot}$ stellar model of solar metallicity \cite{PHW13}. The 
stellar structure was previously calculated using the GENEC stellar evolution code \cite{EMM08}.

By the end of the core He burning phase the $s$-process abundance distribution between 
$^{56}$Fe and $^{100}$Mo was found to be rather insensitive to the MACS values for
$^{23}$Na($n, \gamma$)$^{24}$Na.  Although the MACS at 25 keV (25$-$30 keV is 
the temperature range of the $s$ process during core He$-$burning) is about 10\% lower compared to the previous rate \cite{DPK09}, the final $^{23}$Na overabundance 
increases by only a few \% and the effect on the $s$ abundances between Fe and Sr is limited 
to about 1\%. This is explained by the fact that in He core conditions the $^{23}$Na 
production is marginal, and its abundance coupled with the low MACS implies that the 
neutron poisoning effect of $^{23}$Na during core He burning is low.
  
\begin{figure}
\includegraphics[width=8cm]{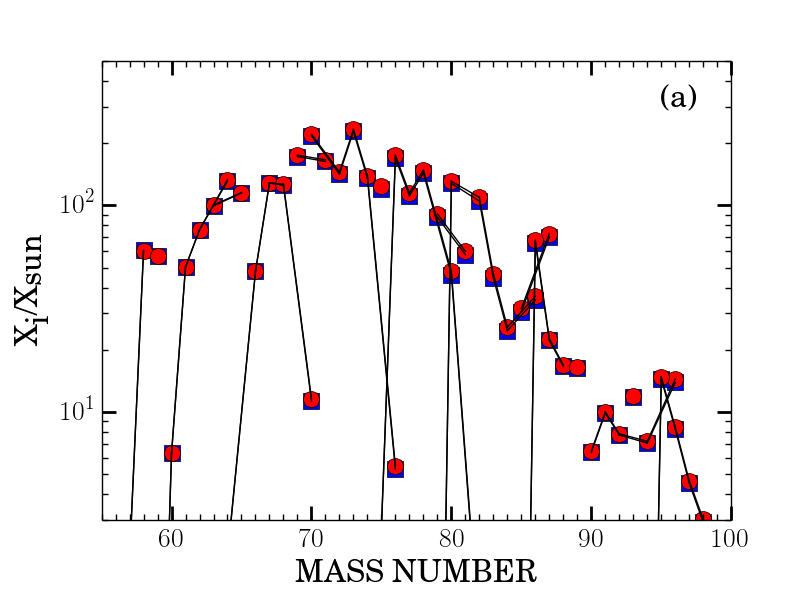}
\includegraphics[width=8cm]{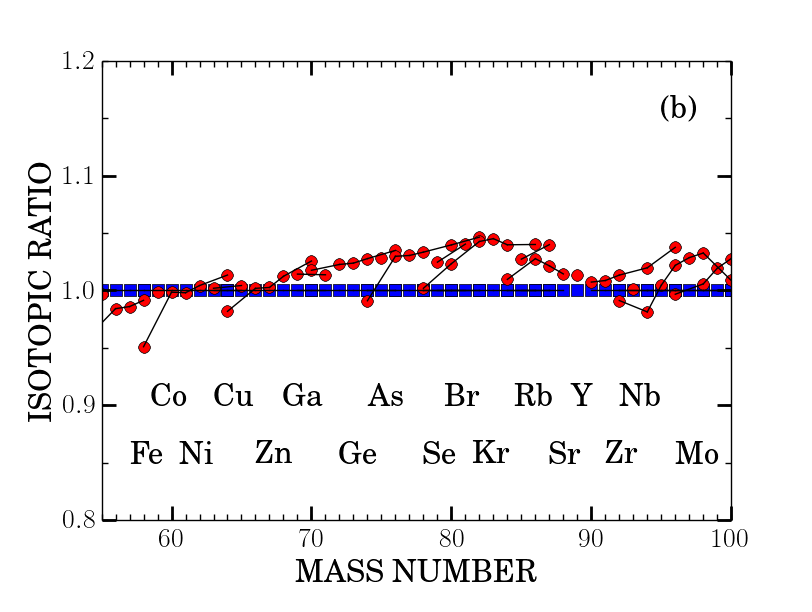}
\caption{Color online. Top: Calculated relative  $s$-abundance distribution at the end of C shell 
            burning for a 25 M$_\odot$ star compared to the distribution obtained with 
           the MACS of $^{23}$Na from the KADoNiS compilation \cite{DPK09}. 
           Bottom:
           Isotopic ratios emphasizing the reduced neutron poison effect due to 
           the smaller MACS of $^{23}$Na from this work. (Isotopes of the same 
	    element are connected by solid lines.)\label{fig3}}
\end{figure}

The final 
$s$-abundance distribution at the end of C shell burning between $^{56}$Fe and $^{100}$Mo 
obtained with the new $^{23}$Na($n, \gamma$)$^{24}$Na MACS is compared in Fig. 
\ref{fig3} with the distribution based on the previous rate \cite{DPK09}. At this point, 
the entire isotopic distribution is affected with variations in the order of 5\%.
At 90 keV thermal energy (typical for the C-burning phase) the new MACS of $^{23}$Na 
is lower by 13\% compared to the previous rate of Bao {\it et al.} \cite{DPK09}. While 
the effect on the final overabundance of $^{23}$Na increases only by about 1\%, the effect 
of $^{23}$Na as an important neutron poison in the C shell becomes evident by the 
propagation effect beyond iron.

Interestingly, the neutron-rich isotopes $^{70}$Zn, $^{76}$Ge, and 
$^{82}$Se, which are traditionally considered to be of $r$-only origin, are  
affected by the new MACS of $^{23}$Na as much as most of the $s$-only isotopes
(e.g., $^{70}$Ge and $^{76}$Se). 

The reduced neutron poison effect of the lower sodium MACS leads to an enhancement 
of the neutron density, thus increasing the neutron capture probability in the $s$-process 
branchings. By far the strongest 
change is obtained for the branching point at $^{79}$Se, where neutron capture on 
the unstable isotope $^{79}$Se becomes more probable. As a consequence, the $^{80}$Kr/$^{82}$Kr 
ratio is reduced by about 3\%. The changes in the $^{85}$Kr branching 
affect mostly the final abundances of $^{86}$Kr and - by the later decay of $^{85}$Kr - 
of $^{85}$Rb, rather than those of the related $s$-only isotopes $^{86,87}$Sr.  Beyond 
the abundance peak around Sr, the $s$-process production in massive stars becomes 
marginal, and the current MACS of $^{23}$Na has a negligible effect.

In the model used in this work the C shell is not convective during the last day before 
the SN. In models, where the C shell stays convective, the neutron density rises from a 
few 10$^{11}$ up to a few 10$^{12}$ because all the residual $^{22}$Ne is 
consumed in ($\alpha, n$) reactions at the final increase of the C-burning temperature 
\cite{PGH10}. In this case, the higher neutron density will lead to a correspondingly 
larger modification of the abundance pattern in the $s$-process branchings.

It is interesting to note that the higher $s$-process efficiency found with the reduced 
MACS data for $^{23}$Na is partly compensated by the effect of revised ($n, \gamma$)
data for the Ne \cite{HPU14} and Mg \cite{MKB12} isotopes. Accordingly, we confirm the 
conclusion of Heil {\it et al.} \cite{HPU14} for the weak component, i.e. that "the 
reproduction of the $s$ abundances in the solar system is far from being settled."
Accordingly, further improvements of the neutron capture cross sections for heavy 
species along the $s$-process path and for light neutron poisons are fundamental 
for constraining $s$-process nucleosynthesis predictions in massive stars. 

\subsection{AGB stars}

There are essentially two mechanisms for sodium production in AGB stars. 
At solar metallicities, sodium is produced primarily during H shell burning 
where the mixing of protons with the He shell gives rise not only to the 
formation of a $^{13}$C pocket (where neutrons are produced via the 
$^{13}$C($\alpha, n$)$^{16}$O reaction), but also to related $^{14}$N and 
$^{23}$Na pockets, thus activating the NeNa cycle in the latter mixing 
zone \cite{Mow99,GoM00}. Under these conditions, neutron 
reactions on sodium are of minor importance.

At low metallicities, however, large amounts of primary $^{22}$Ne are 
synthesized by conversion of primary $^{12}$C into $^{14}$N during 
H burning, which is then transformed during He burning by the sequence 
$^{14}$N($\alpha, \gamma$)$^{18}$F($\beta^+\nu$)$^{18}$O($\alpha, 
\gamma$)$^{22}$Ne \cite{Mow99, GBH06, HGB07}. This $^{22}$Ne 
contributes significantly to the primary production of light isotopes, as 
$^{23}$Na (via $^{22}$Ne($n, \gamma$)$^{23}$Ne($\beta^-$)$^{23}$Na) 
and $^{24}$Mg (via $^{23}$Na($n, \gamma$)$^{24}$Na($\beta^-$)$^{24}$Mg). 
Accordingly, $^{22}$Ne and $^{23}$Na are - together with $^{12}$C, 
$^{14}$N, and $^{16}$O - major neutron poisons in the $^{13}$C 
pocket. As shown in the nucleosynthesis studies of Cristallo {\it et al.} \cite{CSG09}
neutron captures on $^{22}$Ne account for about 50\% of the total sodium 
production at very low metallicity ($Z = 0.0001$). For higher metallicities 
this effect decreases and becomes negligible at [Fe/H]$\geq$$-1$.

With the larger neutron exposures in stars of low metallicity, which are 
characteristic of the strong $s$-process component, the highly
abundant Ne and Na are either acting as seeds for the reaction flow
(enhancing the $s$-process production up to Pb/Bi \cite{HGB07})
or as neutron poisons, depending on the efficiency for neutron 
production in the $^{13}$C pocket.

\begin{figure}
\includegraphics[width=8cm]{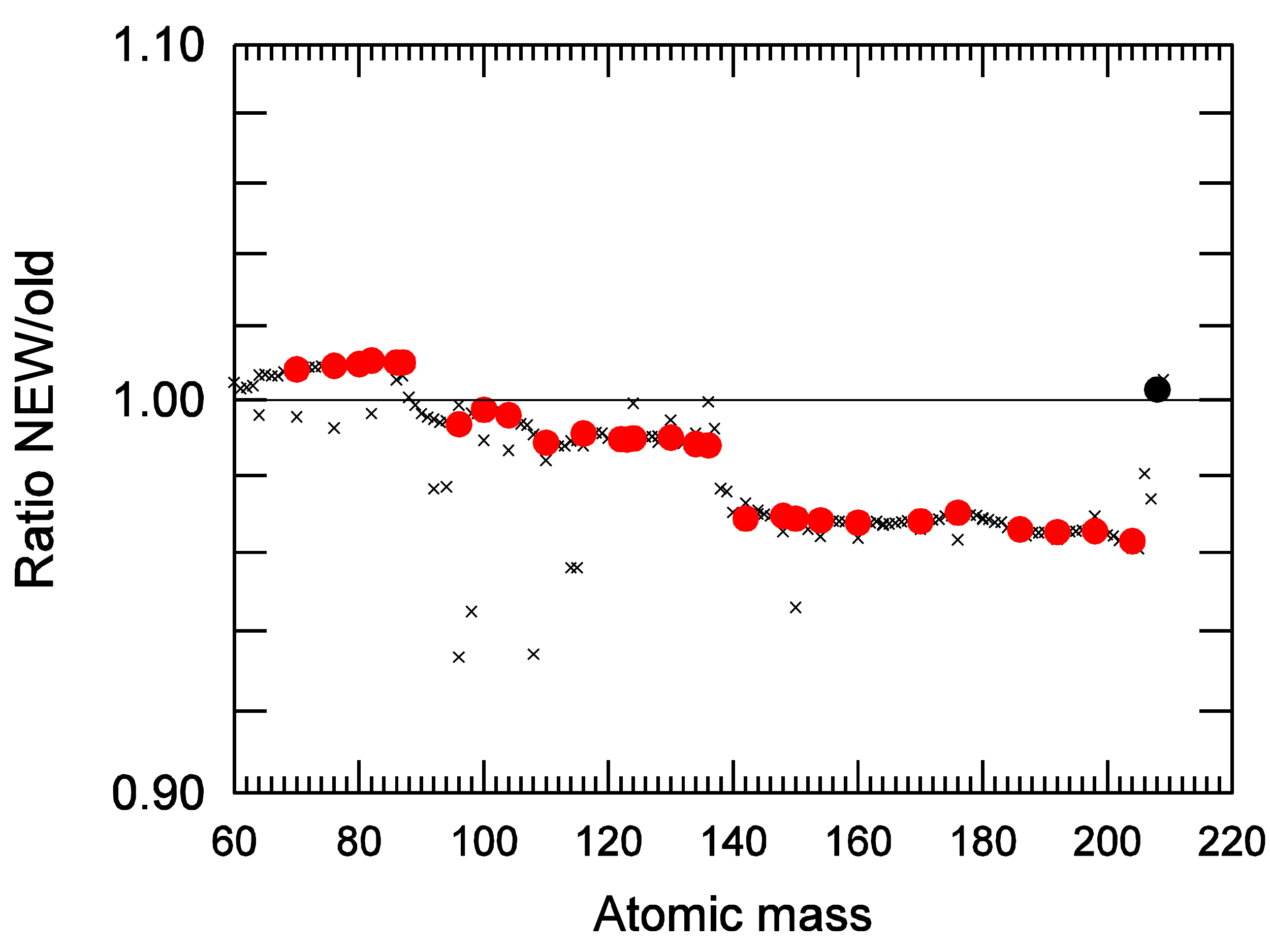}
\includegraphics[width=8cm]{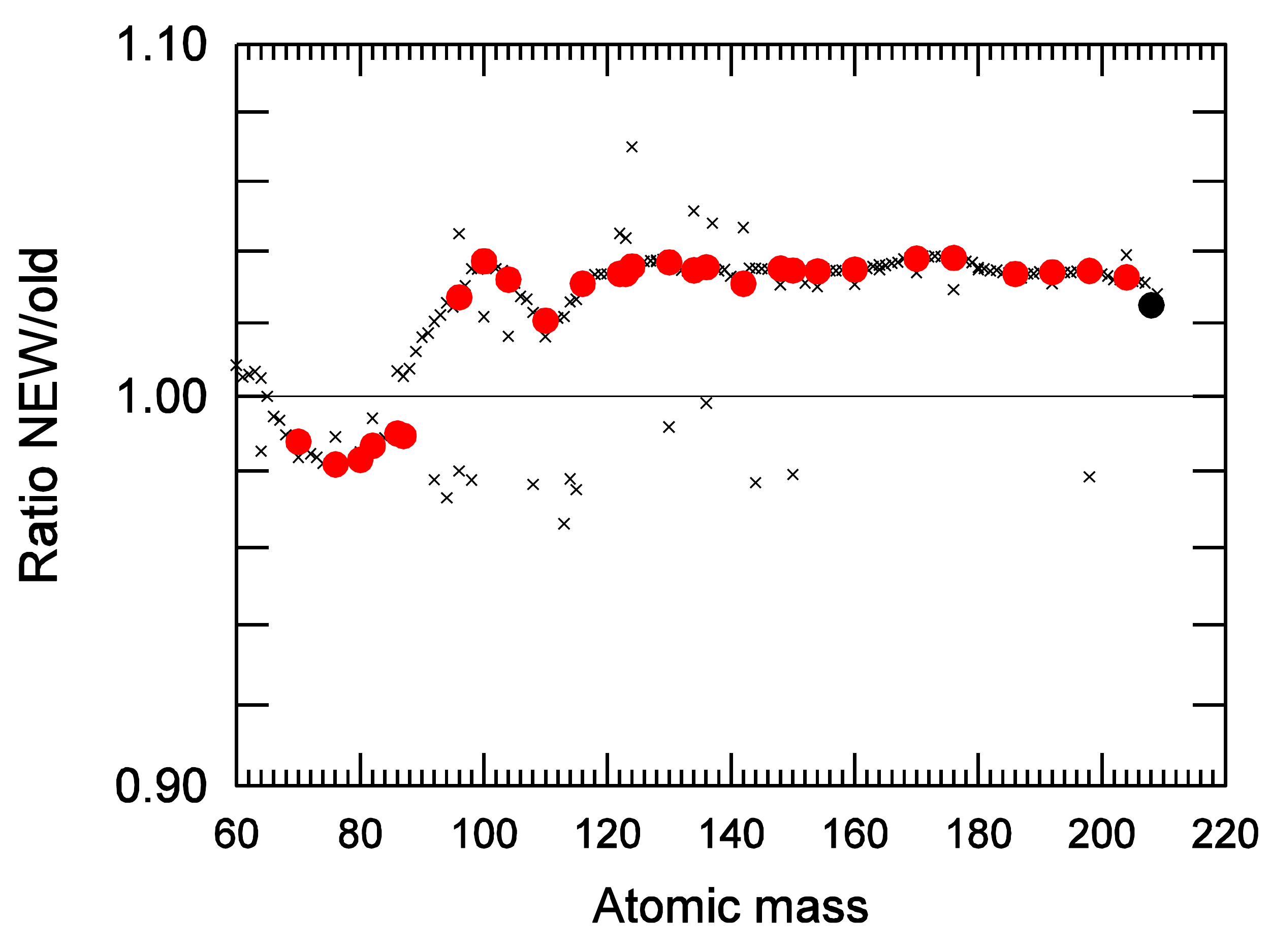}
\caption{Color online. The ratio of $s$-process yields of a 1.5 M$_\odot$ star
with Z=0.0001 obtained with the present and old MACS for $^{23}$Na. 
Pure $s$-nuclei are highlighted by full circles, crosses outside the overall 
distribution are due to branchings in the reaction path. Top: For neutron 
captures in the $^{13}$C pocket $^{23}$Na acts as an additional seed. 
This contribution is reduced by the smaller MACS of this work. Bottom: In 
less efficient $^{13}$C pockets the poisoning effect of $^{23}$Na dominates. 
Consequently, more free neutrons are available due to the smaller MACS, thus 
relaxing the poisoning effect. \label{agb}}
\end{figure}

The effect of the present MACS is illustrated in Fig. \ref{agb} for the 
case of a 1.5 M$_\odot$ star with a $\sim$200 times lower metallicity compared 
to solar (Z=0.0001). For efficient $^{13}$C-pockets (e.g. for the standard 
$^{13}$C pocket (ST) adopted in Ref. \cite{GAB98}) the 
Ne-Na abundances are acting as neutron seeds and are contributing to 
the $s$-process production up to Pb/Bi. With the smaller MACS for 
$^{23}$Na these contributions are reduced, resulting in the relative 
reduction of the $s$-distribution indicated in the upper panel of Fig. 
\ref{agb}. In less efficient $^{13}$C pockets (ST/12) the role of $^{23}$Na 
as a neutron poison becomes dominant as shown in the lower panel. 
Due to the smaller MACS, more free neutrons are now available for the 
$s$-process and are leading to an increase of the $s$-distribution.

One of the major uncertainties of the $s$ process in low-mass AGB stars 
is related to the mixing mechanisms that model the $^{13}$C-pocket.  
A clear answer to the properties involved in such mixing, possibly 
resulting from the interplay between different physical processes in stellar 
interiors (e.g., overshooting, semi-convection, rotation, magnetic fields, see 
review by Herwig \cite{Her05} and Refs. \cite{SGC06,KaL14,NuB14}) has 
not been reached yet, thus leaving the structure of the $^{13}$C-pocket 
a persisting problem.

Depending on the shape and extension of the $^{13}$C pocket, the impact 
of light neutron poisons may affect the $s$ distribution in different ways.
Because the $^{13}$C pocket is artificially introduced in our post-process 
AGB models, the impact of the new $^{23}$Na MACS could be explored
by adopting different shapes and sizes of the $^{13}$C pocket according 
to recent theoretical and observational indications. From the results 
obtained in these tests, the $s$-distribution was affected by less than 
~$\sim$5\%, independent of the assumptions for the $^{13}$C pocket. 
Therefore, the improved accuracy of the present MACS provides 
significant constraints for the neutron poisoning effect of $^{23}$Na in 
AGB stars.

\section{Summary}

The $^{23}$Na($n, \gamma$)$^{24}$Na cross section has been measured at the Karlsruhe 
Van de Graaff accelerator in quasi-stellar thermal neutron spectra at $kT=5.1$ and 25 keV.  
The resulting Maxwellian averaged cross sections of $\langle\sigma\rangle_{\rm kT=5 keV}=
9.1 \pm 0.3$ mb and $\langle\sigma\rangle_{\rm kT=25 keV}=2.03 \pm 0.05$ mb are 
significantly smaller compared to the recommended values of the KADoNiS-v0.3 compilation 
\cite{DPK09}. After reducing the radiative width of the prominent s-wave resonance at 
2.8 keV by 35\%, the measured cross sections were found perfectly compatible with the set 
of resonance data in the ENDF/B-II.1 library. 

With this modification, Maxwellian averaged cross sections in the relevant range of thermal energies between
$kT=5-100$ keV were derived using the energy dependence obtained by an R-matrix 
calculation with the SAMMY code \cite{Lar06}. The effect of the present cross section on the $s$ process
abundances in massive stars (weak $s$ process) is quite small during the He core burning phase,
but becomes significant during the carbon-shell burning phase where $^{23}$Na is synthesized 
in increasing quantities via the $^{12}$C($^{12}{\rm C}, p$)$^{23}$Na reaction. The impact 
of the present MACS measurement has been investigated within a massive star model. It was 
found that the new lower MACS causes a propagation effect over the entire weak $s$-process 
distribution, with a general abundance increase of about 5\%.
\vspace*{5mm}

\begin{center}
{\bf Acknowledgements}
\end{center}

The authors are thankful to D. Roller, E.-P. Knaetsch, W. Seith, and the entire
Van de Graaff group for their support during the measurements.  EU would also 
like to acknowledge the support of JINA (Joint Institute for Nuclear Astrophysics), University 
of Notre Dame, Notre Dame, IN, USA. SB acknowledges financial support from JINA (Joint Institute for 
Nuclear Astrophysics, University of Notre Dame, IN) and KIT (Karlsruhe
Institute of Technology, Karlsruhe, Germany). MP acknowledges support from 
NuGrid by NSF grant PHY 09-22648 (Joint Institute for Nuclear Astrophysics, JINA), 
NSF grant PHY-1430152 (JINA Center for the Evolution of the Elements), and 
EU MIRG-CT-2006-046520.  He also appreciates support from the "Lendulet-2014" 
Programme of the Hungarian Academy of Sciences (Hungary) and from SNF 
(Switzerland). MP also acknowledges PRACE, through its Distributed Extreme Computing Initiative, for resource allocations on Sisu (CSC, Finland), Archer (EPCC, UK), and Beskow (KTH, Sweden) and the support of STFC’s DiRAC High Performance Computing Facilities; DiRAC is part of the National E-infrastructure. Ongoing resource allocations on the University of Hull’s High Performance Computing Facility - viper - are gratefully acknowledged.
CL acknowledges support from the Science and Technology Facilities Council UK (ST/M006085/1).

\newcommand{\noopsort}[1]{} \newcommand{\printfirst}[2]{#1}
  \newcommand{\singleletter}[1]{#1} \newcommand{\swithchargs}[2]{#2#1}

\end{document}